\documentclass{elsart}
\usepackage{graphicx,amssymb}

\begin{document}

\begin{frontmatter}

\title{Dynamical scenario for nonextensive statistical mechanics}

\author{Constantino Tsallis}
\ead{tsallis@cbpf.br}

\address{Centro Brasileiro de Pesquisas Fisicas\\
Rua Xavier Sigaud 150, 22290-180 Rio de Janeiro -- RJ, Brazil \\ 
and\\
Santa Fe Institute \\
1399 Hyde Park Road, Santa Fe, New Mexico 87501, USA 
}

\date{December 10, 2003}

\begin{abstract}
Statistical mechanics can only be ultimately justified in terms of microscopic dynamics (classical, quantum, relativistic, or any other). It is known that Boltzmann-Gibbs statistics is based on the hypothesis of exponential sensitivity to the initial conditions, mixing and ergodicity in Gibbs $\Gamma$-space. What are the corresponding hypothesis for nonextensive statistical mechanics? A scenario for answering such question is advanced, which naturally includes the {\it a priori} determination of the entropic index $q$, as well as its cause and manifestations, for say many-body Hamiltonian systems, in (i) sensitivity to the initial conditions in Gibbs $\Gamma$-space, (ii) relaxation of macroscopic quantities towards their values in anomalous stationary states that differ from the usual thermal equilibrium (e.g., in some classes of metastable or quasi-stationary states), and (iii) energy distribution in the $\Gamma$-space for the same anomalous stationary states.
\end{abstract}


\end{frontmatter}

\section{INTRODUCTION}
There is no reasonable doubt that it was, at least intuitively, neat and clear for Ludwig Boltzmann and for Albert Einstein \cite{einstein,cohenbaranger} that statistical mechanics descends, in one way or another,  from microscopic dynamics. It should be so for all essential concepts of the theory, including the celebrated expression for the entropy, namely (written here for a discrete case)
\begin{equation}
S_{BG} = -k \sum_{i=1}^W p_i \ln p_i \;,
\end{equation}
from now on referred to as the Boltzmann-Gibbs (BG) entropy. However, and interestingly enough, it is until now unknown how to start from say Newton's law $\vec{F}=m\vec{a}$ and end up with this remarkable functional form and its associated variational principle. It is Boltzmann's magnificent intuition which made him to propose this specific microscopic expression for Clausius' thermodynamic entropy. Expression which is totally consistent with Newton's law, although rigorously speaking we still do not know why.

If such is the situation for $S_{BG}$, it is clear that the situation has to be even more intriguing -- to say the least! -- for any generalization of $S_{BG}$, such as \cite{tsallis}
\begin{equation}
S_q = k \frac{1-\sum_{i=1}^W p_i^q}{q-1} \;\;\;\;(q \in \mathbb{R};\; S_1=S_{BG}).  
\end{equation}

Even if we do not yet know how $S_{BG}$ descends from dynamics and what restrictive premises the microscopic dynamics must satisfy, it has been profusely verified that $S_{BG}$ is the natural entropic form everytime we have nonintegrable and sufficiently chaotic dynamics (i.e., {\it positive} Lyapunov exponents), known to yield quick mixing and eventually ergodicity in phase space.  A natural question arises: what happens if the system is nonintegrable but {\it all} Lyapunov exponents vanish? More precisely, what happens when the sensitivity to the initial conditions diverges {\it less} than exponentially? It seems plausible to imagine that the answer depends on {\it how} the sensitivity diverges asymptotically, a power-law?, a slow logarithmic divergence?, some other type? The case that we focus on in the present paper, and generically in nonextensive statistical mechanics, is the ubiquitous {\it power-law}. It is for such dynamics that one imagines that $S_q$ would be the appropriate physical entropy to be used for various purposes, including the bridging to thermodynamics. Furthermore, one imagines that the appropriate value of the entropic index $q$ should reflect the {\it exponent} of the power-law, thus constituting the basis for universality classes of nonextensivity.   

\section{A POSSIBLE TRIANGLE FOR THE ENTROPIC INDEX $q$}

Although most of the notions that we shall use here occur in all types of dynamics, including quantum ones, we shall focus on classical systems for simplicity. 

\subsection{SENSITIVITY TO THE INITIAL CONDITIONS}
Let us start with the notion of {\it sensitivity to the initial conditions}. At the present point it is enough, as an illustration, to consider a one-dimensional nonlinear dynamical system (e.g., an unimodal map like the logistic one) characterized by $x(t)$. We define the sensitivity (to the initial conditions) $\xi \equiv \lim_{\Delta x(0) \to 0}\Delta x(t)/\Delta x(0)$, where $\Delta x(t) \equiv x(t)-x^\prime(t)$ denotes the difference between two trajectories $x(t)$ and $x^\prime(t)$ that are initially close to each other. The following differential equation is typically satisfied:
\begin{equation}
\frac{d \xi}{dt}=\lambda_1 \xi \;,
\end{equation}
where $\lambda_1$ is the Lyapunov exponent (or essentially the largest one in more general situations).  The solution of course is
\begin{equation}
\xi(t)= e^{\lambda_1\,t} \;,
\end{equation}
hence $\ln \xi \propto t$ . 
This solution generically holds everytime $\lambda_1 \ne 0$. But if we have $\lambda_1=0$ (referred to as {\it weak chaos} if there is mixing, and {\it integrability}  otherwise), we expect the adequate differential equation to typically be the following generalization of Eq. (3):
\begin{equation}
\frac{d \xi}{dt}=\lambda_{q_{sen}} \xi^{q_{sen}} \;,
\end{equation}
where {\it sen} stands for {\it sensitivity}. Its solution is given by
\begin{equation}
\xi(t)= e_{q_{sen}}^{\lambda_{q_{sen}}\,t} \;,
\end{equation}
hence $\ln_{q_{sen}} \xi \propto t$ , with \cite{qexp}
\begin{equation}
e_q^x \equiv [1+(1-q)x]^{1/(1-q)} \;\;\;\;(e_1^x=e^x)\;,
\end{equation}
and
\begin{equation}
\ln_q x \equiv \frac{x^{1-q}-1}{1-q} \;\;\;\;(\ln_1 x=\ln x)\;.
\end{equation}
(We remind that $e_q^x$ is always real and nonnegative; for $q<1$, it vanishes for all values of $x$ such that $1+(1-q)x \le 0$). 
 
If $\lambda_1=0$ and, in spite of that, there is mixing in the full dynamical space of the system, we typically expect $q_{sen}<1$ and $\lambda_{q_{sen}}>0$; if there is no mixing, we typically expect $q_{sen}>1$ and  $\lambda_{q_{sen}}<0$. It is of course possible in even more pathological cases (and therefore more rare in natural or artificial systems) that {\it both} $\lambda_1$ and $\lambda_{q_{sen}}$ vanish, but such cases are out of the scope of the present discussion, centered on power-laws (and not on even weaker behaviors such as the logarithmic ones). 

It is by now well known that an illustration of the situation we are focusing  on here is provided by the edge of chaos of the logistic map: $q_{sen}=0.2445...$ and   $\lambda_{q_{sen}}= 1/(1-q_{sen}) = 1.3236...$ \cite{TPZ,fulvioalberto}. Many other paradigmatic dynamical systems are known  or expected to also enter, in one way or another, into the same category. This is the case of globally coupled standard maps close to integrability (\cite{coupledstandard,coupledstandard2,coupledstandard3} and references therein) or long-range-interacting many-body Hamiltonian systems (\cite{longrange} and references therein). 

Let us finally mention that, at least for simple one-dimensional maps, $q_{sen}$ also enters prominently in basic properties such as the entropy production per unit time \cite{catania} (essentially the Kolmogorov-Sinai entropy rate), and the values at which the multifractal function (associated with the edge of chaos) may vanish \cite{lyratsallis}.

\subsection{RELAXATION}
We define the quantity
\begin{equation}
\Omega(t) \equiv \frac{\mathbb{O}(t)-\mathbb{O}(\infty)}{\mathbb{O}(0)-\mathbb{O}(\infty)} \;,
\end{equation}
where $\mathbb{O}$ is a macroscopic observable relaxing towards its value at a stationary state (thermal equilibrium, or some other). The associated differential equation very frequently is
\begin{equation}
\frac{d\Omega}{dt}=-\frac{1}{\tau_1}\,\Omega \,
\end{equation}
$\tau_1>0$ being the relaxation time. The solution of this equation is
\begin{equation}
\Omega(t) =e^{-t/\tau_1} \,.
\end{equation}
The quantity $\tau_1$ is in principle expected to depend on the observable $\mathbb{O}$. However, in all generic cases one expects (according to Krylov's classical remarks) $1/\tau_1 \sim \lambda_1$ if $\lambda_1 >0$. This is quite natural since the relaxation of a macroscopic quantity reflects nothing but the exploration that the system makes of its full  dynamical space in its search of a possible stationary state (given some class of initial conditions), and this occurs at a rythm dictated by the Lyapunov exponent $\lambda_1$ (or, in general, by the full Lyapunov spectrum). In particular, if $\lambda_1$ (or the maximal Lyapunov exponent) vanishes, we expect $\tau_1$ to diverge, thus making Eq. (11) inappropriate.  It is quite plausible that, in such situation, Eq. (10) becomes generalized into
\begin{equation}
\frac{d\Omega}{dt}=-\frac{1}{\tau_{q_{rel}}}\,\Omega^{q_{rel}} \,
\end{equation}
where {\it rel} stands for {\it relaxation}. The solution is of course given by
\begin{equation}
\Omega(t)=e_{q_{rel}}^{-t/\tau_{q_{rel}}} \;\;\;\;(\tau_{q_{rel}}>0)\,
\end{equation}
being typically $q_{rel}>1$. It could well happen that $q_{rel}>1$ depends on the specific observable $\mathbb{O}$ we might be interested in, but even if such is the case, we expect all possible $q_{rel}$'s to be simply inter-related. Moreover, we expect them to be essentially characterized by the speed at which the nonlinear dynamics makes the system to approach its stable (or metastable) stationary (or quasistationary) state, given the class of initial conditions within which the system has been started. More specifically, when $\lambda_1=0$, we expect simple hypothesis such as ergodicity in full phase space (basic for the validity of BG statistical mechanics) to be not fulfilled. Even more, we expect (under conditions to be determined) the system to have a tendency of lengthily living in some kind of attractor or pseudo-attractor, {\it whose Lebesgue measure in full dynamical space would typically be zero}, as it is the case of (multi)fractals or scale-free networks. If we assume an ensemble whose Lebesgue measure is different from zero at $t=0$ inside a special region of full space, we might observe a shrinking of the Lebesgue measure as time goes on. The loss of Lebesgue measure would possibly occur through a $q$-exponential function whose value of $q$ would be noted $q_{rel}$. From now on, we reserve the notation $q_{rel}$ for the shrinking of Lebesgue measure, if such is the case. All other $q_{rel}$'s would be simply related to this one. 
Let us emphasize that the present scenario does {\it not} exclude that, at even larger times, the system escapes from this (multi)fractal ``prison" and ultimately satifies ergodicity, thus making a crossover from $q_{rel}$ to $q=1$. This would in fact be the case unless some limiting situations (see \cite{coupledstandard2}) are taken which strictly guarantee that $\lambda_1$ vanishes. When such {\it escape} is possible, {\it aging} phenomena might be present \cite{montemurro,robledoaging}, aging being presumably related to a slow collective dynamics which would prepare the escape from the ``prison".  

A typical illustration of the picture we have described in the present subsection is, as before, the edge of chaos of the logistic map, where it is $q_{rel} \simeq 2.4$ \cite{lyra}. 
Other paradigmatic illustrations should include the systems mentioned in the previous subsection, but this remains to be verified.

Let us finally mention that, at least for simple one-dimensional maps, $q_{rel}$ also emerges \cite{lyra} in basic properties such as the fractal dimension (maximum of the multifractal function) of the attractor at the edge of chaos.

\subsection{STATIONARY STATE}

Let us now address what may in some sense be considered as the most useful thermostatistical function, namely the distribution of total energies in Gibbs $\Gamma$-space of a thermodynamical system in contact with a canonical thermostat fixing the temperature $T$. 

Let us first consider a Hamiltonian system whose elements interact only {\it locally}, or almost {\it not at all} (e.g., the ideal gas). A typical example of local interaction is
a classical two-body (attractive) potential which has no singularity at the origin, and which asymptotically decays with distance $r$ as $ -1/r^\alpha$ with $\alpha/d>1$, where $d$ is the space dimension of the system; for instance, for the usual Lennard-Jones fluid, we have $\alpha=6$ and $d=3$. We know that the relevant stationary state is thermal equilibrium and that
\begin{equation}
p_i = \frac{e^{-\beta_1 E_i}}{Z_1}   \;\;\;\;(\beta_1 \equiv 1/kT; Z_1 \equiv \sum_{j}e^{-\beta_1 E_j}) \;,
\end{equation}    
where $\{E_i\}$ is the set of eigenvalues of the Hamiltonian with specific boundary conditions. 
This celebrated distribution can be obtained through the textbook procedure which consists in optimizing $S_{BG}$ under the constraints $\sum_ip_i=1$ and $\sum_ip_iE_i=U_1$, where $U_1$ is the total internal energy. It seems to be realized by only a few physicists that this distribution can {\it also} be obtained, {\it albeit} without formal justification up to now, through a differential equation, namely
\begin{equation}
\frac{d(p_i Z_1)}{d E_i}=-\beta_1 (p_i Z_1)\;.
\end{equation}
The black-body radiation law was first found, and published in October 1900 \cite{planck}, by Max Planck through an heuristic differential equation, precisely one which, with appropriate variables, contains Eq. (15) as a particular case! This is known only by those interested in the history of science. And it has not been incorporated in standard textbooks presumably because of the lack of formal justification. Nevertheless the mathematical fact is there, and in my opinion it is perhaps not a mere coincidence. Indeed, since the sensitivity to the initial conditions follows an exponential law, and the same typically does its prominent consequence namely the relaxation of macroscopic thermostatistical quantities, it cannot be a very great surprise that the same does happen to the distribution associated with the stationary state towards which the system relaxes.

For Hamiltonian systems whose elements do not interact locally but {\it globally} (e.g., the classical systems mentioned previously but with $0 \le \alpha/d \le 1$, in which case the potential is {\it not} integrable), it seems quite plausible that the Eq. (15) becomes generalized into
\begin{equation}
\frac{d(p_i Z_{q_{stat}})}{d E_i}=-\beta_{q_{stat}} (p_i Z_{q_{stat}})^{q_{stat}}\;,
\end{equation}
where {\it stat} stands for {\it stationary state}, and $\beta_{q_{stat}}$ is the adequate inverse temperature (in units of $k$). The solution is given by
\begin{equation}
p_i = \frac{e_{q_{stat}}^{-\beta_{q_{stat}} E_i}}{Z_{q_{stat}}}   \;\;\;\;(\beta_{q_{stat}} \equiv 1/kT_{{stat}}; Z_{q_{stat}} \equiv \sum_{j}e_{q_{stat}}^{-\beta_{q_{stat}} E_j}) \;,
\end{equation}  
and one typically expects $q_{stat}>1$. It is in fact {\it precisely} this solution which emerges \cite{tsallis} by optimizing, through conveniently written norm and energy constraints, $S_q$ instead of $S_{BG}$.

There is molecular dynamical numerical evidence \cite{evidence} for some classical long-range-interacting many-body inertial rotators on $d$-dimensional lattices that they might satisfy,  in some robust metastable states in the presence of a thermostat, the total energy distribution indicated in Eq. (17). But, even at the numerical level, the problem is uneasy, though not impossible, to study. This is the subject of ongoing efforts. The strategy which is followed is to consider an isolated system of $N$ rotators at a fixed total energy. Then consider once for ever a subset of $M$ rotators among the $N$ available. Finally, one must numerically approach the $t \to \infty$ limit of the $M \to \infty$ limit of the $N \to \infty$ limit. The whole procedure must be repeated many times ($N_{realizations} \to \infty$, mathematically speaking) for realizations differing in the precise multi-particle initial conditions, always within some specific and nontrivial class of initial conditions, and then averaging the results. The distribution obtained in this manner (i.e., $\lim_{t \to \infty} \lim_{M \to \infty} \lim_{N \to \infty} \lim_{N_{realizations} \to \infty}$) is to be compared with Eq. (17). 

Let us make one more remark, which concerns the distributions of momenta of these rotators. If we denote with $ \mathbb{H}(\{L_i\},\{\theta_i\})$ the Hamiltonian associated with say planar rotators, the distribution in $\Gamma$-space is given by
\begin{equation}
p(\{L_i\},\{\theta_i\}) \propto e_{q_{stat}}^{-\beta_{q_{stat}}\mathbb{H}(\{L_i\},\{\theta_i\})} \;.
\end{equation}
Then the one-body angular momentum marginal distribution $p(L)$ is given by
\begin{eqnarray}
p(L) &=& \int (\Pi_{i=2}^N\, dL_i)   \, (\Pi_{i=1}^N\, d\theta_i)  \,    p(\{L_i\},\{\theta_i\})         \nonumber  \\       
&\propto &              \int (\Pi_{i=2}^N \, dL_i)   \, (\Pi_{i=1}^N\, d\theta_i)  \,  e_{q_{stat}}^{-\beta_{q_{stat}}\mathbb{H}(\{L_i\},\{\theta_i\})} \;.
\end{eqnarray}
We do not yet know what is the result for this distribution but it could also be a $q_L$-exponential in $L^2$ with $q_L >1$ related to $q_{stat}$ is some still unknown manner. One expects (essentially because the integral of a $q$-exponential is another $q$-exponential) the index $q_L$ to either coincide with $q_{stat}$ or be a simple function of it (see \cite{mendestsallis}). In any case, the available numerical evidence is not inconsistent with such posibilities (\cite{coupledstandard2} and references therein). See, for instance, Fig. 1 (from \cite{coupledstandard2}), where for the first time the momentum distribution is calculated (through molecular dynamics) for a system which is sensibly closer to a {\it canonical} ensemble (contact with a large thermostat) than the {\it microcanonical} ensemble considered up to now in the literature for such systems.

\begin{figure}
\begin{center}
\includegraphics[width=9.5cm,angle=0]{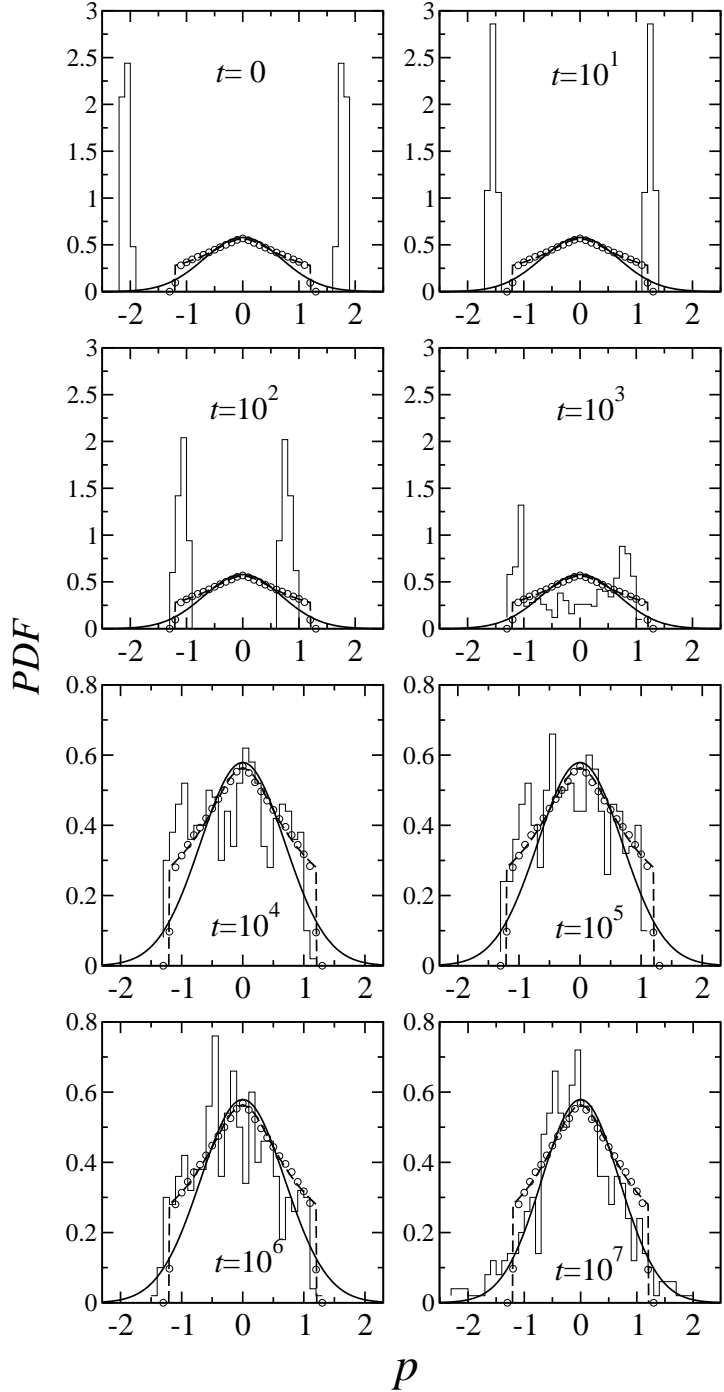}
\end{center}
\caption{\small Snapshots of the evolution of PDF
for momenta in the $(L,\theta)-space $ (i.e., in $\mu$-space),
starting with a ``hot water bag''. In solid black
circles, the instantaneous PDF at time $t=10^k$
(with $k=0,1,...7$), for $M=500$ spins inside an $N=5000$ spin
system. In empty circles, the average of $10^3$ realizations
for $N=10^5$ spins (all averages made during the $QSS$
plateau). In dashed line, the $q_L$-exponential fitting curve
with $T=T_{QSS}=0.38$ and $q_L=3.7$. Finally, in solid line, the
analytical Gaussian PDF. These last three curves are the
same in every frame, and are plotted for reference. From \cite{coupledstandard2}.}
\label{fig_HMF_PDF}
\end{figure}

\begin{figure}
\begin{center}
\includegraphics[width=12cm,angle=0]{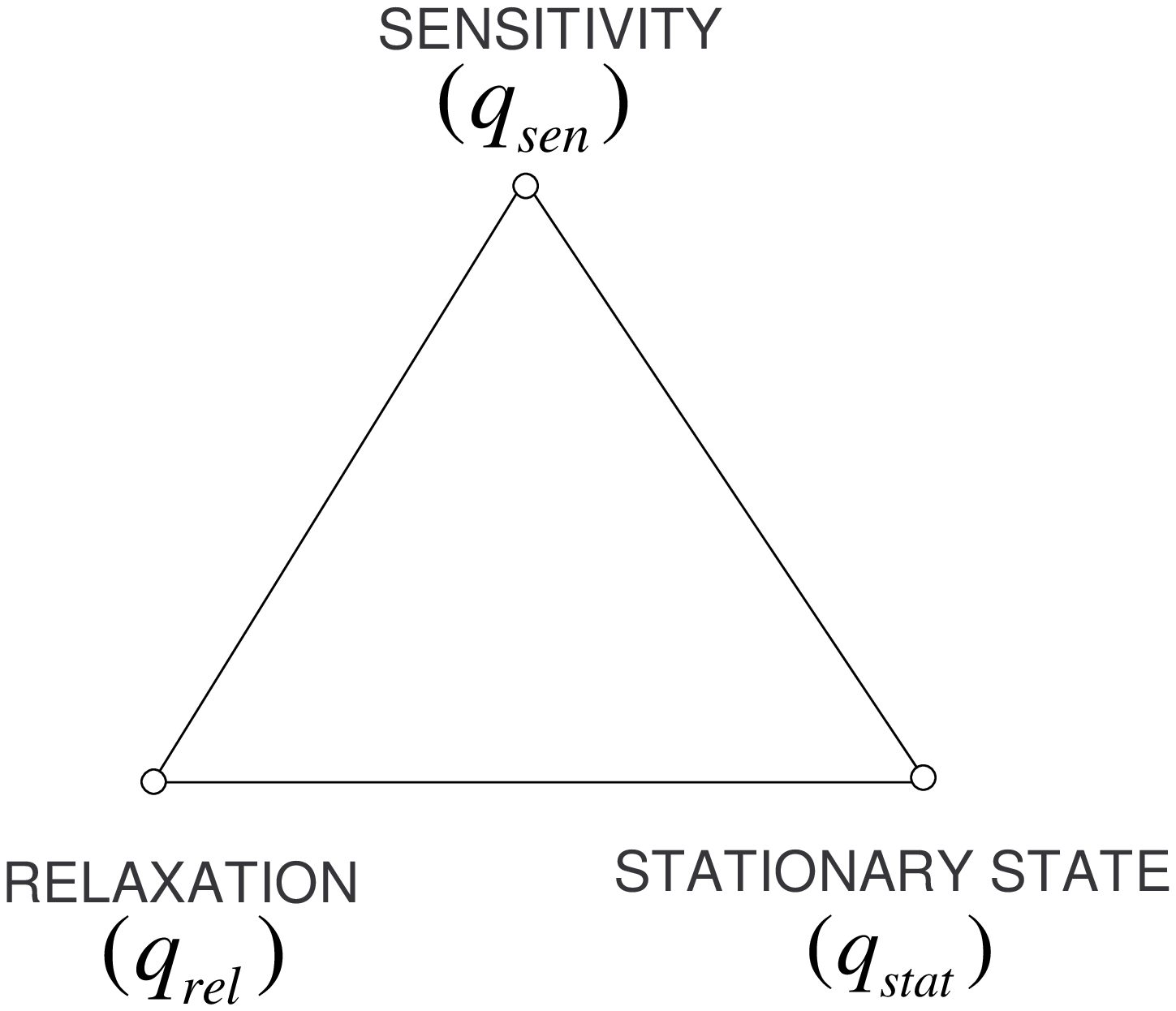}
\end{center}
\caption{\small 
The triangle of the basic values of $q$, namely those associated with sensitivity to the initial conditions, relaxation and stationary state. For the most relevant situations we expect $q_{sen} \le 1$, $q_{rel} \ge 1$ and $q_{stat} \ge 1$. 
These indices are presumably inter-related since they all descend from the particular dynamical exploration that the system does of its full phase space. For example, for long-range Hamiltonian systems characterized by the decay exponent $\alpha$ and the dimension $d$, it could be that $q_{stat}$ decreases from a value above unity (e.g., $2$ or $3/2$) to unity when $\alpha/d$ increases from zero to unity. For such systems one expects relations like the (particularly simple) $q_{stat}=q_{rel}=2-q_{sen}$ or similar ones. In any case, it is clear that, for $\alpha/d>1$ (i.e., when BG statistics is known to be the correct one), one has $q_{stat}=q_{rel}=q_{sen}=1$. All the weakly chaotic systems focused on here are expected to have well defined values for $q_{sen}$ and $q_{rel}$, but only those associated with a Hamiltonian are expected to {\it also} have a well defined value for $q_{stat}$. 
}
\label{fig_triangle}
\end{figure}

\section{FINAL REMARKS}

The $q$-indices discussed in Section 2 and their possible interconnections are schematically represented in Fig. 2. It is our scenario that they are different ``faces" of the same basic phenomenon, namely how the system ``likes" to evolve and mix in part or all of its full phase space (Gibbs $\Gamma$-space for Hamiltonian systems) given its initial conditions. If the system is strongly chaotic (i.e., positive Lyapunov exponents), essentially satisfying Boltzmann's ``molecular chaos hypothesis", then the system will be mixing and ergodic allover the entire phase space. BG statistical mechanics is expected to appropriately describe the thermal statistics of the system, the associated microscopic expression for the entropy undoubtedly being $S_{BG}$. But if the system is only weakly chaotic (i.e., zero Lyapunov exponents, but {\it not} integrable), then the occupancy of phase space might be much more complex everytime the system is started within specific nonzero-Lebesgue-measure regions of phase space. The region of phase space where the system may lengthily evolve before eventually occupying the entire space might have a (multi)fractal or hierarchical structure of the type currently referred to as scale-free network (see \cite{barabasi} and references therein). As long as the system remains in that region, its thermostatistics might be adequately be described within nonextensive statistical mechanics, the associated microscopic expression for the thermodynamic entropy presumably being $S_q$ with a specific value of $q$. During the metastable or quasi-stationary state we would then have $(q_{sen},q_{rel},q_{stat}) \ne (1,1,1)$, and at later times (in many cases, infinitely later!) a crossover would occur to $q_{stat}=q_{rel}=q_{sen}=1$. This confluence onto a single value of $q$ for BG statistics might be at the origin of occasional confusion within the  interested community. Indeed, three essentially different (though related) concepts just happen to coincide within the BG formalism.

\section*{ACKNOWLEDGMENTS}
It is a deep pleasure to thank the creators and organizers of this wonderful meeting in Cagliari (Villasimius) where we all received exceptionally warm hospitality. Although they were many and very valuable, it is for me but a matter of elementary justice to gratefully name those who contributed the most: P. Quarati, G. Kaniadakis, A. Rapisarda, M. Lissia and E.G.D. Cohen.  

Partial support by FAPERJ, CNPq and
PRONEX (Brazilian agencies) is gratefully acknowledged.

\end{document}